\documentclass[preprint]{elsarticle}

\usepackage{hyperref}
\usepackage{amsmath,amssymb}
\usepackage{siunitx}
\usepackage{caption}
\usepackage{subcaption}
\usepackage{fourier} 
\usepackage{charter}

\usepackage{soul}

\biboptions{sort&compress}
\newcommand{\JJ}{\boldsymbol{J}}
\newcommand{\nn}{\boldsymbol{n}}

\newcommand{\sss}{\boldsymbol{s}}

\newcommand{\xx}{\boldsymbol{x}}

\newcommand{\Reff}{R_{\textrm{eff}}}
\newcommand{\nrel}{n_{\textrm{rel}}}
\newcommand{\figwidth}{0.6}

\makeatletter
\def\ps@pprintTitle{%
    \let\@oddhead\@empty
    \let\@evenhead\@empty
    \def\@oddfoot{\footnotesize\itshape
         {Submitted preprint} \hfill\today}%
    \let\@evenfoot\@oddfoot
    }
\makeatother

\begin{document}

\begin{frontmatter}

\title{Fresnel reflection boundary for radiative transport lattice Boltzmann methods in highly scattering volume}

\author[MVM,LBRG]{Albert Mink\corref{IKE}}
\cortext[IKE]{Corresponding author}
\ead{albert.mink@kit.edu}
\author[BVT]{Kira Schediwy}
\author[LBRG]{Marc Haussmann}
\author[BVT]{Clemens Posten}
\author[MVM]{Hermann Nirschl}
\author[MVM,LBRG,MATH]{Mathias J. Krause}

\address[MVM]{Institute for Mechanical Process Engineering and Mechanics, Karlsruhe Institute of Technology, Germany}
\address[BVT]{Institute of Process Engineering in Life Sciences, Karlsruhe Institute of Technology, Germany}
\address[LBRG]{Lattice Boltzmann Research Group, Karlsruhe Institute of Technology, Germany}
\address[MATH]{Institute for Applied and Numerical Mathematics, Karlsruhe Institute of Technology, Germany}


\begin{abstract}
With its roots in kinetic theory, the lattice Boltzmann method (LBM) cannot only be used to solve complex fluid flows but also radiative transport in volume.
The present work derives a novel Fresnel boundary scheme for radiative transport LBM, based on Fresnel's equation, which depicts the partly reflected radiation on surfaces.
Driven from a boundary modeling and discussion on the microscopic level, incorporating Fresnel's equation, it is developed a boundary model for the mesoscopic radiative transport LBM.
At an intermediate step, the Fresnel's equation is related to well known partial differential (Robin) equations, based on a bottom-up approach where the $P_1$-Approximation is deployed.
To connect the novel boundary scheme to the so derived target equation, a Chapman--Enskog expansion is examined in addition.
Both techniques together, point out how to interpret microscopic modeling by the means of macroscopic expressions and as a consequence how, to chose simulation parameters according to the specific boundary.
The numerical tests suggest that the proposed boundary is first order convergent.
The paper closes with a showcase, where the novel boundary method for radiative transport LBM is applied to a setup with multiple LED spots.
\end{abstract}


\end{frontmatter}


\section{Introduction}
In recent years, lattice Boltzmann methods~(LBM) have been established as an alternative to traditional computational fluid dynamics solvers.
Its particle based formulation offers a natural and intuitive modeling of physical phenomena and attracts a wide spectrum of researchers.
In addition, the resulting algorithm is embarrassingly parallel which suits perfectly to the current trend in increasing parallelism in computing hardware~\cite{mohrhard:19}.
Under these circumstances the ongoing effort of extending LBM to a vast range of scientific engineering issues~\cite{aidun:10, krueger:17, succi:18} is not surprising.

The pioneer work in solving the radiative transport equation~(RTE) with an LBM based algorithm has been presented by Geist et al.~\citep{geist:04} in 2004.
This work rather aimed to render real-time photo-realistic pictures, than providing a methodical approach to radiative transport lattice Boltzmann methods (RTLBM).
Some years later, researchers started to derive RTLBM systematically through discretization of RTE, to mainly solve thermal radiation problems in two dimension~\citep{asinari:10, mishra:14, bindra:12, ma:11}.
Simultaneously, a new field of application for RTLBM has been developed, the simulation of radiative transport in participating media or volume~\citep{mcHardy:16, mink:16, yi:16, zhang:13}.
RTLBM is a powerful numerical method for simulation of radiative transport in volume, that can be further extended to transient radiation~\citep{zhang:13, zhang:15, guo:16, wang:17, gairola:17}, to polarized radiation~\citep{zhang:16} and to the visible light spectrum~\citep{mcHardy:18}.
However, the conventional assumption in literature of a purely transmitting or absorbing boundary is not capable to model general engineering applications, in particular when the boundaries are close to the light source.
Therefore radiative transport LBM is still limited to academic examples and lacks on generality.

This work aims to provide a physically based boundary modeling for radiative transport LBM.
To this goal, radiation interaction with surfaces is investigated on a microscopic or photon level, which can be thought as the boundaries of the radiative transport domain.
The radiation that hits a surface can be classified in refracted (transmission), absorbed (absorptance) or reflected (reflection) radiation.
However, it is sufficient to concentrate on the modeling of the refracted part, since the main task of LBM boundary formulation is to find the unknown incoming particles, here radiation.

The present work models the partly reflected radiation by the Fresnel's equation.
Unfortunately, this equation alone cannot  plugged into a LBM straight away, but we propose a two step strategy to find a suitable LBM formulation.
In a first step, which assumes a highly scattering volume, the Fresnel's equation is approximated by a macroscopic equation, examining the $P_1$-Approximation.
In the second step, it is examined a Chapman--Enskog expansion, with the purpose to fully connect the LBM boundary approach to the approximated target equation from the present approximation.
The basic idea of this strategy is, to clearly understand of the modeling parameter by means of macroscopic terms and then develop a suitable LBM boundary scheme.
This ensures the crucial linkage of modeling and simulation parameter and allows to apply the scheme to general boundaries in radiative transport applications.

The mayor finding of the work is the partial bounce-back scheme, that covers the partly reflected radiance on the boundary, according to the Fresnel's equation.
From a numerical standpoint, this scheme inherits the benefits of standard bounce-back boundaries in LBM, most significantly the simple implementation and robustness.
The attraction of the performed Chapman--Enskog analysis along the $P_1$-Approximation, lies in clearly connect and understand the modeling parameter by means of simulation parameter.
This allows the simulation of general boundaries for radiative transport problems with an LBM.

\begin{figure}[ht]
  \centering
  \includegraphics[width=\figwidth\linewidth]{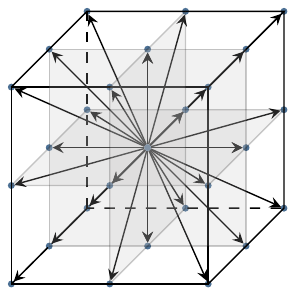}
  \caption{A discrete LBM stencil with $27$ directions (blue dots), also known as $D3Q27$ stencil.
           }
  \label{fig:lbmCube}
\end{figure}

Commonly, lattice Boltzmann methods are formulated on the mesoscopic level, which is a stochastic abstraction of the purely particle based microscopic level.
This idea is borrowed from the kinetic theory, where a large number of (gas) molecules, electrically charged particles or photons, are expressed in terms of particle density distribution functions.
The law of large numbers ensures the asymptotic accuracy of such stochastic concepts.
The present work models the configuration of photons by a mesoscopic distribution function~$f(t,\xx,\sss)$, which describes the amount of density propagating into direction~$\sss$ at time~$t$ and position~$\xx$.
The LBM typical domain discretization results in a regular grid, where in addition every grid-point, also known as cell, is equipped with a set discrete directions~$\sss_i$, see Fig.~\ref{fig:lbmCube}.
The domain discretization is hence trivial and the grid can be easily divided into sub-grids for parallel execution on high performance computers.

The heart of every LBM is the alternating sequence of the collision and streaming step.
The first relaxes a discrete particle density function~$f_i$, associated to direction~$\sss_i$, towards an equilibrium state~$f_i^{eq}$.
Followed, by an streaming step, that propagates the post-collision state~$f_i^\star$ to the neighbouring cell in direction~$\sss_i$.
Both steps read
\begin{align*}
    f_i^\star (t,x) &= f_i(t,x) - \frac{1}{\tau} \left( f_i(t,x) - f_i^{eq}(t,x) \right)\;, \\
    f_i(t+1,x+s_i) &= f_i^\star (t,x)\; ,
\end{align*}
for relaxation time~$\tau$ in which~$f_i$ relaxes towards~$f_i^{eq}$.
The so computed distribution functions~$f_i$ are further associated to macroscopic quantities through averaging the directions.
Here, the radiant energy density
\begin{equation*}
u(t,\xx) = \sum_i f_i(t,\xx)\;,
\end{equation*}
is obtained by averaging the local distributions.

The essential difficulty in any application of LBM, lies in verifying that the simple collision and streaming step, solves asymptotically the desired partial differential equation.
To clarify this issue, researches examine a multi-scale analysis such as the Chapman--Enskog expansion and prove that the correct asymptotic behaviour is achieved.

The remainder of the paper is organized as follows.
The next section reviews the basic quantities in the topic of radiation, along the radiative transport in volume and the Fresnel's equation.
Section~\ref{sec:numericalMethod} presents the radiative transport LBM and the novel boundary scheme.
Finally, the results of numerical investigation are shown and discussed in section~\ref{sec:results}.

\section{Model development}
This section introduces the modeling of radiation and is mainly based on the following textbooks~\citep{wang:07, modest:13}.

\subsection{Modeling radiation}
The present work models radiation as small, virtual particles that propagate on trajectories through space.
Depending on the wavelength~$\lambda$, these particles carry the \emph{photon energy} or \emph{radiant energy}
\begin{equation*}
E(\lambda)
=
\frac{\hslash \, c} {\lambda}
\quad
\quad
\text{in~}
\si{\joule}
\;,
\end{equation*}
with \emph{Planck constant}~$\hslash$ given in~\si{\joule\second}, constant speed of light in vacuum~$c$ in~\si{\meter\per\second} and specific wavelength~$\lambda$ in~\si{\meter}.
Associating the photon energy to a certain time interval the \emph{photon power} is derived by $P(t,\xx) = dE/dt$ at position $\xx$.

\begin{figure}[ht]
  \centering
  \includegraphics[width=\figwidth\linewidth]{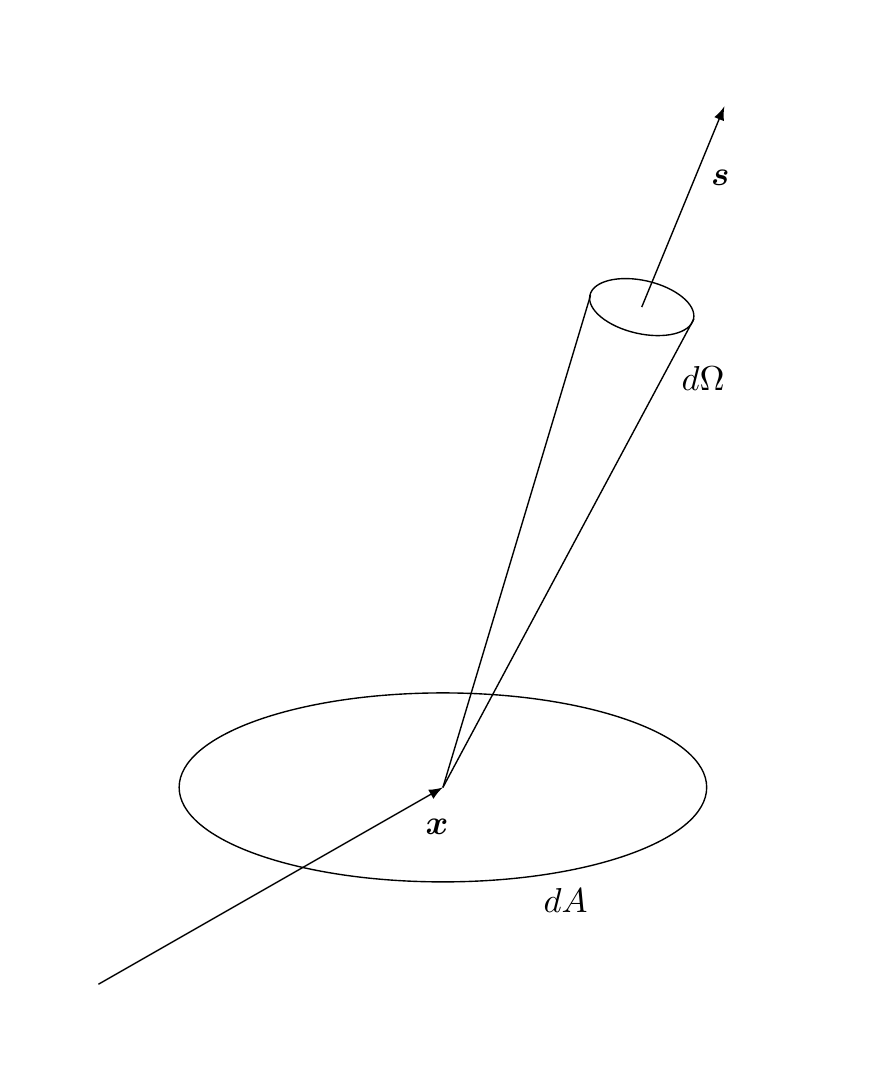}
  \caption{For any fixed time $t$ expression $L(t,\xx,\sss) \, d\textrm{A} \, d\Omega$ describes the incident photon power from direction~$d\Omega$ to surface element~$d\textrm{A}.$
           }
  \label{fig:radiance}
\end{figure}

By taking the propagation direction into account, the \emph{radiance} is defined, as the radiant power per unit normal area and per unit solid angle
\begin{equation*}
L(t,\xx,\sss)
=
\frac{d^2 P(t,\xx)} {d\textrm{A} \, d\Omega}
\quad
\quad
\text{in~}
\si{\joule\per\second}
\times
\si[per-mode = fraction]{\per\square\meter}
\times
\si[per-mode = fraction]{\per\steradian}
\;,
\end{equation*}
with respect to time~$t$, position~$\xx$ and into direction~$\sss$.
The abstract quantity radiance can be thought of, as the amount of photons that enter a cone oriented in the direction of $\sss$, see Fig.~\ref{fig:radiance}.

\subsection{Radiative transport in volume}
Radiance propagating through optically interacting volume, e.g. milk, biological tissue or clouds, is not only governed by the passed distance, but absorption and scattering behavior.
Let us considering radiance that hits a single droplet or volume element from direction~$\sss$.
Then, the radiance diminishes due to out-scattering in an arbitrary direction and absorption.
Conversely, there might be radiance associated to a traveling direction~$\sss'$, that equally gets out-scattered into direction~$\sss$, which then yields in a gain of the radiance in direction~$\sss$, named in-scattering.
Taken together these phenomena, the \emph{radiative transport equation} (RTE)
\begin{equation*}
\frac{1}{c} \, \partial_t L
+
\sss \cdot \nabla L
=
-
(\sigma_a+\sigma_s) \, L
+
\sigma_s
\int_{4\pi}
  p(\sss',\sss) \,
  L(t,\xx,\sss')
\; d\Omega'
\;,
\end{equation*}
describes the radiance~$L(t,\xx,\sss)$ in a volume with specific absorption and scattering coefficients $\sigma_a$ and $\sigma_s$ and phase function~$p$.
The speed of radiation is given by~$c$.

The first term on the right hand side describes the loss of radiance propagating in direction~$\sss$, according to absorption and out-scattering.
The second term accounts for the gain of radiance due to in-scattering and is mainly responsible for the numerical difficulties in solving general radiative transport problems.
The probability of incident radiance from $\sss$ being scattered into direction $\sss'$, is denoted by $p(\sss,\sss')$.
Unlike isotropic scattering volume, that results in a constant phase function, anisotropic scattering behavior, such as Mie or Rayleigh scattering, can be covered by the corresponding phase function.
For further reading it is referred to the textbook Modest~\citep{modest:13}.

The radiance on the particle level can than be linked to macroscopic observable quantities by integration.
This procedure results in the \emph{radiant energy density}\footnote{
  Note that the integral is stated in spherical coordinates and reads
  $
  \int_{4\pi}
  L(t,\xx,\sss) \,
  d\Omega
  =
  \int_{\theta=0}^\pi
  \int_{\phi=0}^{2\pi}
    L(t,\xx,\sss) \sin\theta \,
  d\theta
  d\phi
  ,
  $
  for
  $
  \sss =
  \left(
  \sin\theta \cos\phi,
  \sin\theta \sin\phi,
  \cos\theta
  \right)^\intercal
  .
  $
}
\begin{equation}
\label{eq:radiantEnetryDensity}
u(t,\xx)
=
\frac{1}{c}
\int_{4\pi}
  L(t,\xx,\sss) \,
  d\Omega
\quad
\quad
\text{in~}
\si{\joule\per\cubic\meter}
\;,
\end{equation}
for the speed of radiation~$c$ and can be interpreted as the brightness or photon density of a volume element.

Besides, the radiant energy density, there is a further important macroscopic quantity the \emph{radiant flux} defined by
\begin{equation*}
\JJ (t,\xx)
=
\int_{4\pi}
  \sss \, L(t,\xx,\sss) \,
  d\Omega
\quad
\quad
\text{in~}
\si{\joule\per\second}
\times
\si[per-mode = fraction]{\per\square\meter}
\;,
\end{equation*}
which expresses the averaged propagation direction of the radiance.
Now, that it is clear how to model radiative transport in volume, it is investigated the boundary layer of such a volume.

\subsection{Radiative transport on surface}
The case of radiance interacting with surfaces defines the boundary effects for general radiative transport problems, e.g. radiance passing from water into glass.
The boundary layer is in between both volumes and is characterized by the refractive indices $\nn_i$ and $\nn_o$, for the inner volume, respective outer volume.
For our interest, there are two main modeling aspects.
At its simplest case, the radiance leaves the volume of interest or in other words, there is no radiance re-entering the volume.
Such a boundary can be written, for all boundary points~$\xx$, in the form
\begin{equation*}
\int_{\sss\cdot\nn >0} L(t,\xx,\sss) \; d\Omega
=
0\;
\end{equation*}
where expression $\sss\cdot\nn >0$ ensures that only the directions that pointing into the volume are taken into account.
This phenomena is named transmission and can be observed for two volumes with the same refractive index.

In the general case, radiance is partly reflected on the interface and re-enters the volume.
The portion of the re-entering radiance is described by \emph{Fresnel's equation}.
Since the key part of LBM boundary modeling is to find appropriate incoming particle density function on the boundary, the present work discusses the partly reflected portion on the surface.

\begin{figure}[ht]
  \centering
  \includegraphics[width=\figwidth\linewidth]{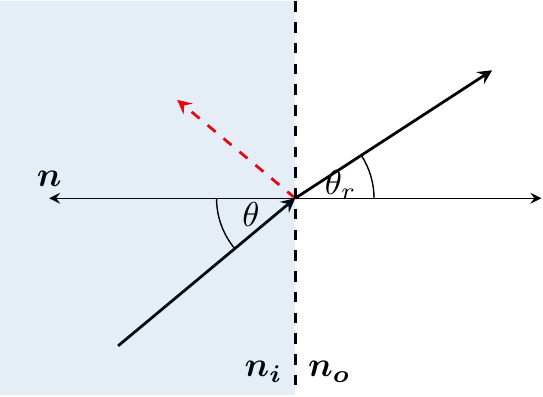}
  \caption{Dashed line represents the interface of two volumes with refractive index~$n_i$ and~$n_o$.
           Incident radiance from angle~$\theta$ changes its direction to angle~$\theta_r$ when passing into the second volume according to Snell's law.
           Dashed arrow (red) indicates the reflected portion governed by Fresnel's equation.
}
  \label{fig:snell}
\end{figure}

For a in-depth understanding of the interaction of radiance with surfaces, it is provided a brief recall of Snell's law, illustrated in Fig.~\ref{fig:snell}.
Given a surface normal~$\nn$ and volume specific refractive indices~$n_i$, $n_o$ with $n_i < n_o$, Snell's law states the change in the propagation direction by the \emph{refraction angle}
\begin{equation*}
\theta_r (\theta)
=
\arcsin
\left( \frac{n_i}{n_o} \sin\theta \right)
,
\end{equation*}
for an incident angle~$\theta$.
In the case of~$n_i > n_o$, a critical angle is defined as
$\theta_c
=
\arcsin\left( \frac{n_o}{n_i} \right)
$
, where total reflection is observed for~$\theta > \theta_c$ and radiance is entirely reflected (mirrored at~$\nn$) back into medium.

\begin{figure}[ht]
  \centering
  \includegraphics[width=\figwidth\linewidth]{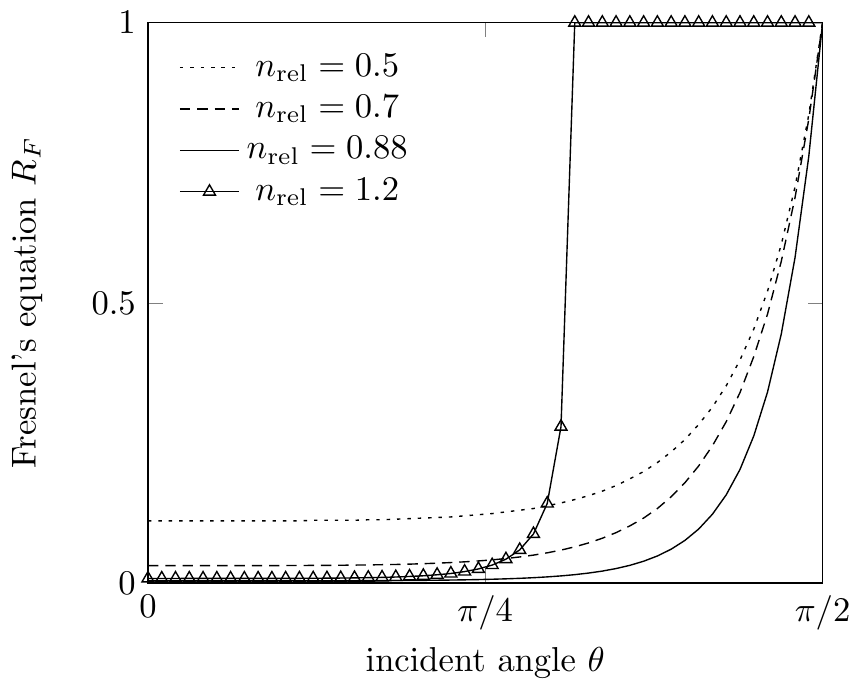}
  \caption{Fresnel's equation $R_F$ plotted over incident angle $\theta$ for varying relative refractive index $\nrel$.
           Total reflection is observed for a water-air interface ($\nrel=1.2$) only for radiation that hits the interface almost parallel.
           Other directions are almost fully transmitted and pass the boundary interface.
           }
  \label{fig:reflectionfunction}
\end{figure}

The key modeling equation of the present work, the \emph{Fresnel's equation}, reads
\begin{equation}
\label{eq:fresnelrelation}
R_F(\theta)
=
\left\{
  \begin{array}{l}
  1, \quad \text{for } \theta > \theta_c,
  \\
  \frac{1}{2}
  \left(
    \frac{\nrel \cos\theta_r -\cos\theta}
         {\nrel \cos\theta_r +\cos\theta}
  \right)^2
  +
  \frac{1}{2}
  \left(
    \frac{\nrel \cos\theta -\cos\theta_r}
         {\nrel \cos\theta +\cos\theta_r}
  \right)^2, \text{ for } 0\leq \theta \leq \theta_c.
  \end{array}
\right.
\end{equation}
Given the boundary specific \emph{relative refractive index}~$\nrel = n_i/n_o$, the Fresnel's equation associates an incident angle to a certain reflectivity in the interval~$[0,1]$.
The special case of $R_F=1$ models total reflection, e.g. mirrors, and $R_F=0$ pure transmission, e.g. $n_i = n_o$.
An illustration of the strong dependence of $R_F$ on the incident direction is shown in Figure~\ref{fig:reflectionfunction}.
For the sake of compactness, it is written $\theta_r$ instead of $\theta_r(\theta)$.

Assuming that radiance hits a interface at position $\xx$ from direction $\theta$.
The re-entering or partly reflected radiance can be stated informally by
\begin{equation}
\label{eq:FresnelBoundary}
 L(t,\xx,\theta') = L(t,\xx,\theta) R_F(\theta) \;,
\end{equation}
for an reflected angle $\theta_r$, compare Fig.~\ref{fig:snell}.
According to this formula, the reflected radiance (left hand side) is the product of incident radiance and Fresnel's equation (right hand side).
As a first guess, one could derive an LBM scheme from~\eqref{eq:FresnelBoundary}, by choosing a large set of discrete directions to properly resolve the angular dependency and reduce the discretization error.
Then however, every cell is equipped with this large set of directions, which installs an undesired coupling of boundary discretization error and numerical costs.

This downside can be avoided by, first substituting~\eqref{eq:FresnelBoundary} by a less direction-sensitive boundary formulation and second, deriving a discrete version.
To this aim, there are two quantities introduced to average the direction of radiance.
First, the \emph{irradiance} is introduced
\begin{equation*}
\JJ_{\textrm{ir}} (t,\xx)
=
\int_{\nn \cdot \sss > 0}
  L(\sss)\,
  \sss \cdot \nn \,
  d\Omega
\quad
\quad
\text{in~}
\si{\joule\per\second}
\times
\si[per-mode = fraction]{\per\square\meter}\;,
\end{equation*}
which accounts for the total incident radiance on a surface point with orientation~$\nn$.
Accordingly, the average outgoing radiance is introduced by the \emph{radiosity}
\begin{equation*}
\JJ_{\textrm{ra}} (t,\xx)
=
\int_{\nn \cdot \sss < 0}
  L(\sss)\,
  \sss \cdot \nn \,
  d\Omega
\quad
\quad
\text{in~}
\si{\joule\per\second}
\times
\si[per-mode = fraction]{\per\square\meter}
\;.
\end{equation*}
Both quantities are associated to a position~$\xx$ with the surface normal~$\nn$, where incident directions satisfy $\nn \cdot \sss > 0$ and contrary holds for outgoing directions.
By expanding the radiance in the basis of spherical harmonics, up to first order, it yields
\begin{equation*}
L(t,\xx,\sss) = \frac{1}{4\pi} u(t,\xx) + \frac{3}{4\pi} \JJ(t,\xx) \cdot \sss
\;,
\end{equation*}
and subsequently the irradiance is reformulated, by means of $u$ and $\JJ$, to
\begin{equation}
\label{eq:irrad}
\JJ_{\textrm{ir}} = \frac{1}{4} u + \frac{1}{2} \JJ \cdot \nn
\;.
\end{equation}
This is a reasonable approximation for the radiance, if highly scattering volumes are considered.
With these findings, the averaged version of~\eqref{eq:FresnelBoundary} is derived in the following.

\subsection{\texorpdfstring{$P_1$-}{}Approximation for Fresnel boundary}
\label{ssec:p1}
This section follows the concept of averaging the direction and derives a macroscopic formula for the Fresnel boundary.
By summarizing the reflected portion of all incident radiance on a boundary surface element, equation~\eqref{eq:FresnelBoundary} reads
\begin{equation}
\label{eq:mesoscopicrobin}
\int_{\nn\cdot\sss > 0}
  L(\sss)\,
  \sss \cdot \nn \,
d\Omega
=
\int_{\nn\cdot\sss < 0}
  R_F(\sss\cdot\nn)\,
  L(\sss) \,
  \sss \cdot \nn \,
d\Omega
\;.
\end{equation}
This integral equation can be thought of as the overall re-entering radiance on the boundary (left hand side), equals the (overall) reflected part of the incident radiance, given by the right hand side.
The argument $(t,\xx)$ has been dropped for the sake of compactness.

In the following, the above boundary equation is approximated by means of macroscopic quantities and under the assumption of a highly scattering volume.
The effective reflection coefficient is introduced by
\begin{equation*}
\Reff
=
\frac{
\int_{\sss\cdot\nn <0}
  R_F(\sss\cdot\nn)\,
  L(\sss) \,
  \sss \cdot \nn \,
  d\Omega
}
{
\int_{\sss\cdot\nn <0}
  L(\sss) \sss \cdot \nn \, d\Omega
}
\;.
\end{equation*}
With~\eqref{eq:irrad} and the definitions
\begin{align*}
R_u
&=
\int_0^{\pi/2}
  2\, \sin\theta\; R_F(\theta) \, d\theta
\;,
\\
R_{\JJ}
&=
\int_0^{\pi/2}
  3\, \sin\theta\;\left(\cos\theta \right)^2 R_F(\theta) \, d\theta
\;
,
\end{align*}
the effective reflection coefficient is retrieved
\begin{equation}
\label{eq:reff}
\Reff
=
\frac{R_u + R_{\JJ}}
{2-R_u+R_{\JJ}}
\;.
\end{equation}
The details of this purely algebraic operations can be found in~\cite{wang:07}.
What is achieved by now, is that~$\Reff$ is no longer expressed in terms of the unknown radiance and the incident direction, but can be computed solely by means of~$R_F$ which is equivalent to knowing~$\nrel$, see definition~\eqref{eq:fresnelrelation}.

In the next step is directly concluded from~\cite{wang:07} and the Robin boundary formulation is obtained.
Starting of with the substitution of~\eqref{eq:reff} into~\eqref{eq:mesoscopicrobin}, it yields
\begin{equation*}
\int_{\nn\cdot\sss > 0}
  L(\sss)\,
  \sss \cdot \nn \,
d\Omega
=
\Reff
\int_{\nn\cdot\sss < 0}
  L(\sss) \,
  \sss \cdot \nn \,
d\Omega
\;.
\end{equation*}
Followed by plugging-in the approximated irradiance and radiosity~\eqref{eq:irrad}, we get
\begin{equation*}
\frac{1}{4} u + \frac{1}{2} \JJ \cdot \nn
=
\frac{1}{4} \Reff\, u - \frac{1}{2} \Reff\, \JJ \cdot \nn
\;,
\end{equation*}
and finally, with Fick's law ($\JJ = -D\nabla u $) and the following definition
\begin{equation}
\label{eq:CR}
C_R
=
\frac{1+\Reff}{1-\Reff}
\;,
\end{equation}
the macroscopic Robin boundary condition
\begin{equation}
\label{eq:robin}
u(t,\xx)
-
2 \, C_R D \nabla u(t,\xx) \cdot \boldsymbol{n}
=
0
\;,
\end{equation}
is recovered.
Here, $D = 1/(3(\sigma_a+\sigma_s))$ is a diffusion coefficient and $C_R$ a constructively derived dimensionless parameter accounting for the surface reflection.
The Robin boundary is a sum of Dirichlet and Neumann boundary, where product $C_R D$ indicates whether the Dirichlet or Neumann part dominates.
For example, Neumann part~$\nabla u$ becomes dominant for~$C_R D \to \infty$ and contrary for~$C_R D \to 0$ condition~$u = 0$ dominates.

Table~\ref{tab:cr} shows typical values for $C_R$ (water-glass $\nrel=0.88$).
The reflection parameter $C_R$ depends on the relative refractive index~$\nrel$ through the definition of~$\Reff$ and hence is the boundary surface specific parameter.
The volume is represented in the boundary equation by the diffusion parameter.
For refractive index match, $n_i = n_o$, the derived macroscopic equation does not depend on the reflection parameter~$C_R$, which is in perfect agreement to the underlying Fresnel's equation.
\begin{table}[ht] 
\centering
\begin{tabular}{l||l|l|l|l}
  $\nrel$ & 1.0 & 0.88 & 0.7 & 0.5 \\   \hline
  $C_R$     & 1.0 & 1.0478 & 1.14 & 1.35
\end{tabular}
\caption{Refractive index match ($\nrel=1$) results in $C_R =1$.
         Parameter~$C_R$ increases for relative refractive index $\nrel<1$.}
\label{tab:cr}
\end{table}

\section{Numerical method}
\label{sec:numericalMethod}
After introducing the radiative transport LBM for volume, this section presents the novel Fresnel boundary.
The proposed LBM scheme is analyzed by a Chapman--Enskog expansion and shown to fulfill the macroscopic formulation.

\subsection{The radiative transport lattice Boltzmann method}
Considering photons at a specific wavelength\footnote{
  The specific wavelength is the equivalent of the monatomic gas assumption in modeling fluid flow.
  Remember that in standard LBM mass density~$f$ is a function in~
  \si{\kg} $\times$
  \si[per-mode=reciprocal]{\per\cubic\meter} $\times$
  \si[per-mode=fraction]{(\meter\per\second)^{-3}}.
}
, the \emph{photon density distribution function}
reads
\begin{equation*}
f(t,\xx,\sss)
\quad
\quad
\text{in~}
\si{\joule}
\times
\si[per-mode = fraction]{\per\cubic\meter}
\times
\si[per-mode = fraction]{\per\steradian}
\;.
\end{equation*}
For any infinitesimal volume element~$dx \, d\Omega$, the expression $f(t,\xx,\sss) \, dx \, d\Omega$ denotes the local photon density in direction~$\sss$ at position~$\xx$ with respect to a given time~$t$.

The corresponding \emph{discrete photon density} is defined as
\begin{equation*}
f_i(t,\xx)
=
w_i \, f(t,\xx,\sss_i)
\;,
\end{equation*}
for a given LBM stencil~$DnQq$, that comes with the definition of discrete directions~$\sss_1, \ldots, \sss_q$, positive weights~$w_1, \ldots, w_q$ and dimension~$n$.
These stencils are chosen such, that the radiant energy density~\eqref{eq:radiantEnetryDensity} is computed exactly by a Gauss--Hermite quadrature formula
\begin{equation*}
u(t,\xx) = \sum_{j=1}^q f_j(t,\xx)
\;.
\end{equation*}

The present work deploys the $D3Q7$\footnote{
  For $D3Q7$ grids the weights are given by
  $w_1 = 1/4, w_2 = \ldots = w_7 = 1/8$ and the discrete directions by
  $\sss_1 = (0,0,0)^T $,
  $\sss_2 = (-1,0,0)^T $,
  $\sss_3 = (0,-1,0)^T $,
  $\sss_4 = (0,0,-1)^T $,
  $\sss_5 = (1,0,0)^T $,
  $\sss_6 = (0,1,0)^T $,
  $\sss_7 = (0,0,1)^T $.
}
stencil and is based on the second order RTLBM derived in Mink et al.~\citep{mink:16}.
In order to solve the steady state radiative transport in volume, they proposed the stream and collide equation
\begin{equation}
\label{eq:rtlbm}
f_i(t+1, \xx + \sss_i )
=
f_i
-
\frac{1}{\tau^\star} \, ( f_i - f_i^{eq} )
-
\frac{\sigma^\star_a}{D^\star} \, \frac{f_i}{8}\
\;,
\end{equation}
for an equilibrium function~$f_i^{eq}(u)= w_i u$, non-dimensional absorption parameter~$\sigma^\star_a$ and diffusion coefficient~$D^\star$ and \emph{relaxation time}~$\tau^\star$ (non-dimensional).
The latter covers the time evolution of the system and can be set to one in RTLBM~\citep{wolf:95, patil:14}, since steady state of radiative transport is reached instantaneously.
In~\eqref{eq:rtlbm} the following standard abbreviations $f_i = f_i(t,\xx)$ and $f_i^{eq} = f_i^{eq}(u)(t,\xx)$ are applied.

\subsection{Boundary treatments}
Boundaries for radiative transport might be non-permeable, which stands out from many fluid flow problems.
The reason for that is, that radiation can either be transmitted, emitted or reflected on the boundary, even a combination of all microscopic phenomena.

\subsubsection{Neumann and Dirichlet boundary}
One of the most elementary and fundamental boundary formulation in LBM is the bounce\hbox{-}back scheme, where the outgoing particle~$f_i$ re\hbox{-}enters\footnote
{
  In LBM literature bounce\hbox{-}back is commonly described by reflection of direction.
  In order to avoid abuse of language, we use reflect in microscopic modeling and re-enter for mesoscopic context.
}
in the opposite direction~$\bar i$.
One possible implementation consists of replacing~\eqref{eq:rtlbm} on the boundary by
\begin{equation*}
    f_i(t+1,\xx+\sss_i)
    =
    f_{\bar{i}}(t,\xx),
\end{equation*}
where $\bar{i}$ is the opposite direction of $i$.
For RTLBM this scheme corresponds to a zero Neumann boundary for the radiant energy density, since there is no change in radiance on a boundary cell.
This work applies the full way bounce\hbox{-}back implementation and subsequently the collision on the boundary is realized by a trivial swap of~$f_i$ by~$f_{\bar{i}}$, see the original work on bounce-back for LBM~\citep{cornubert:91, ziegler:93}.

An additional boundary model is the Dirichlet boundary that fixes the density on the boundary at a constant value.
Mink et al.~\cite{mink:16} applied the Dirichlet boundary to radiative transport LBM, which is shown to be in excellent agreement to analytic solution.

A common boundary for the radiative source is to fix the inward pointing~$f_i$, such that the expected flux is enforced.
This boundary cam be found in McHardy et al.~\cite{mcHardy:16}.

\subsubsection{Partial bounce-back}
Fresnel's equation states that outgoing radiance is partially reflected and re-enters the computational domain and the corresponding Fresnel boundary equation is formulated in~\eqref{eq:FresnelBoundary}, respectively~\eqref{eq:mesoscopicrobin}.
The present work follows the microscopic picture and proposes a \emph{partial bounce-back} scheme
\begin{equation}
\label{eq:mesorobin}
f_{\bar i}(t+1,\xx_b+\sss_i)
=
r_F \,
f_i (t, \xx_b)
\;
\end{equation}
in order to solve the Fresnel boundary equation, for a wet boundary node~$\xx_b$, an outgoing direction~$i$ and its opposite direction~$\bar{i}$.
Based on derivation in~\ref{sec:ce}, the \emph{mesoscopic reflection function} is found in the form
\begin{equation}
r_F
=
1
-
\frac{2}
{4 C_R D^\star +1}
\;,
\end{equation}
for non-dimensional diffusion coefficient~$D^\star$ and~$C_R$ as in~\eqref{eq:CR}.
In other words, the Fresnel's equation is reduced in the simulation to a sole reflectivity of the wall given by~$r_F$, the LBM counterpart of~$R_F$.

The innovation of this boundary model lies in a precise and physically based description of the mesoscopic reflection function.
Besides that, it is equipped with a clear and simple implementation of the relevant parameter, namely~$C_R$ that covers the Fresnel's boundary at the surface and additionally the volume properties that are represented by~$D$.
As a bonus, the algorithm is almost as simple as a standard bounce-back scheme and therefore very well suited for high performance computing.

\begin{figure}[ht]
  \centering
  \includegraphics[width=\figwidth\linewidth]{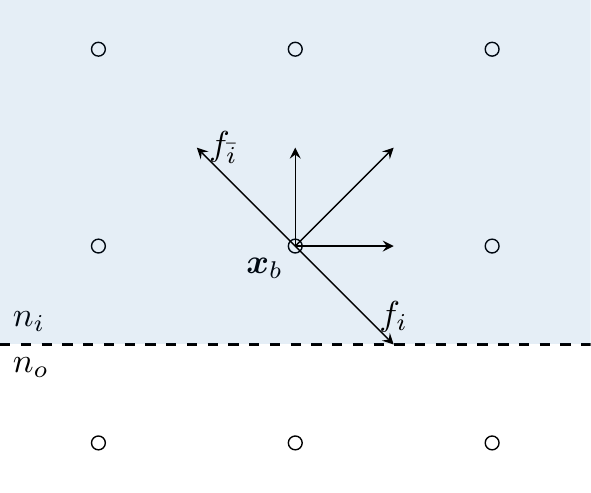}
  \caption{Boundary region represented by the dashed line with a wet boundary node~$\xx_b$.
           Particle~$f_i$ hitting the boundary surface re-enters after collision on the opposite direction~$\bar{i}$.
           The change of the re-entering photon density~$f_{\bar{i}}$ is given by mesoscopic reflection function~$r_F$.
           }
  \label{fig:partialbb}
\end{figure}

For the quantitative considerations, it should be noted that the mesoscopic reflection function~$r_F$ takes values in the interval~$[-1,1]$.
In its most simple case, $r_F =1$, the boundary scheme is equivalent to bounce\hbox{-}back, thus photon density~$f_i(t, \xx_b)$ re-enters totally at the opposite direction~$\bar i$, see Fig.~\ref{fig:partialbb} and simulation results in Fig.~\ref{fig:threshold}.
This is in excellent agreement to macroscopic considerations, see section~\ref{ssec:p1}, where for $D^\star C_R \to \infty$ the zero Neumann boundary for the density is obtained.
In fact, the limit of $D^\star C_R \to \infty$ result in $r_F \to 1$ and hence the partial bounce-back scheme fulfills the macroscopic prediction.

\section{Results}
\label{sec:results}
\sisetup{per-mode=reciprocal}
To validate the novel Fresnel boundary model, numerical simulations have been carried out.
All simulations are implemented in \emph{OpenLB}\footnote{\url{www.openlb.net}}, an open-source LBM solver, with respect to $D3Q7$ stencils and the novel boundary implementation will be published in the upcoming release.

\subsection{Validation Setup}
\begin{figure}[ht]
  \centering
  \includegraphics[width=\figwidth\linewidth]{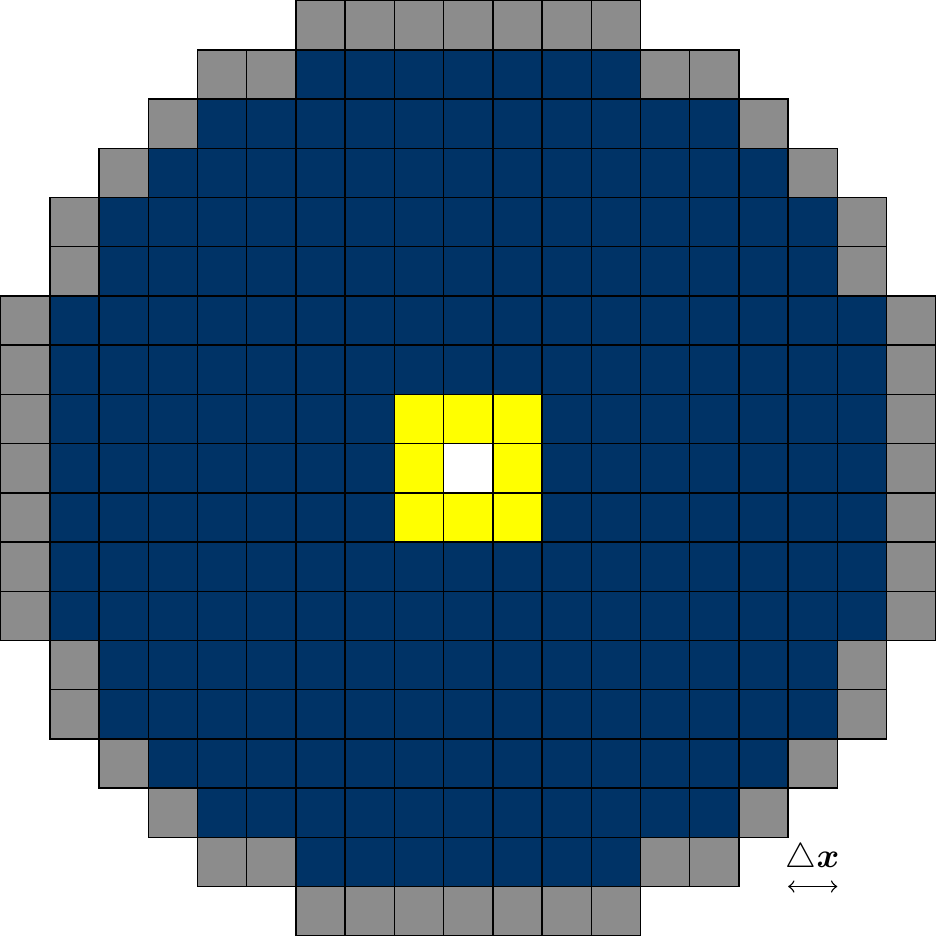}
  \caption{Cross-section of three dimensional discrete geometry, with cell size $\triangle x$.
           Yellow indicates light source, blue is participating volume and grey is the discussed boundary.
           White voxels are neither boundary nor computational domain, hence outside the computational domain.
           }
  \label{fig:simulationsetup}
\end{figure}

The validation of the proposed RTLBM boundary~\eqref{eq:mesorobin} is based on a spherical geometry setup, with a spherical light source inside, see cross-section in Fig.~\ref{fig:simulationsetup}.
Here, the light source is placed at a distance~\SI{0.1}{\meter} from the center and the investigated boundary at a fixed distance~\SI{1}{\meter}.
For a given resolution~$N$, the regular grid is resolved by cells of size~$\triangle x:= N^{-1}$ in~\si{\meter}.
This simulation setup is a natural choice, since for Dirichlet boundaries at the light source and the novel boundary at the outlet, there are known extrapolated distances against which we validate.

Unless otherwise stated, absorption and scattering is given by
$\sigma_a = \SI{0.5}{\per\meter}$
and
$\sigma_s = \SI{1.5}{\per\meter}.$
Further, the refractive index for the participating volume is given by
$n_i = 1.33$ (water) and for boundary by
$n_o = 1.51$ (glass).
Given the above parameters, the macroscopic refraction coefficient results in $C_R=1.0478$, see Table~\ref{tab:cr} and diffusion coefficient in~$D=\SI{1/6}{\meter}$.
The corresponding non-dimensional versions are $D^\star = D\triangle x^{-1}$, $\sigma_a^\star = \sigma_a \triangle x$ and $\sigma_s^\star = \sigma_s \triangle x$.

Steady state criteria for simulations is a density deviation in~$L^2$-norm of less than~\num{1e-5} during a period of~$ N^2 $ iteration steps.
The time scale is proportional to $(\triangle x)^2$ and thus diffusive scaling is deployed.
Finally, the light source is defined as Dirichlet radiant energy density, accordingly to the previous work~\citep{mink:16}.




\subsection{Validation of the partial bounce-back boundary}
\label{sec:validation}
\begin{figure}[ht]
  \centering
  \includegraphics[width=\figwidth\linewidth]{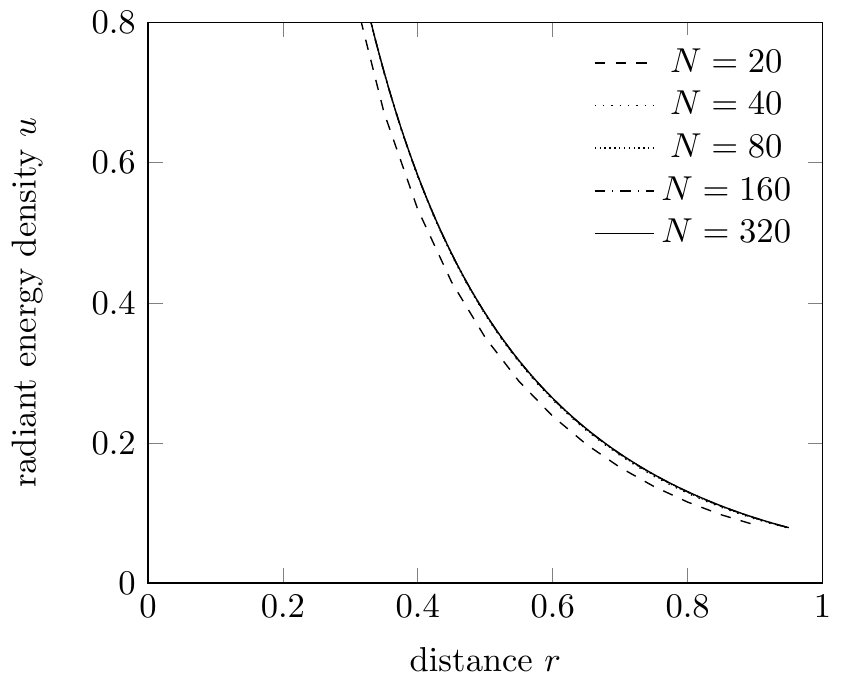}
  \caption{Simulated radiant energy density $u$ over distance to origin $r:=|x|$.
           Grid convergence for the proposed algorithm~\eqref{eq:rtlbm} and Robin boundary~\eqref{eq:mesorobin} is observed.
           }
  \label{fig:gridindependent}
\end{figure}

First, numerical simulations have been carried out to prove that the algorithm is stated grid independently and converges.
For this purpose, the simulation data along the positive x-axis is investigated, which is a reasonable choice due to the symmetry of computational domain.
In Figure~\ref{fig:gridindependent} simulations for varying resolutions $N$ and fixed absorption and scattering coefficients are shown.
It is clearly seen that a resolution of $N=40$ or $\triangle x =\SI{0.025}{\meter}$ is already enough to resolve the problem properly and higher resolutions do not yield in substantially more accurate simulations.
This observation suggests, that the algorithm is grid convergent, including the novel boundary formulation.
At the resolution of~\num{40} the steady state is reached after about~\num{10500} iterations.
Finally, the iterations can be dramatically reduced if the domain is initialized with an appropriate guess, for example the solution known for Dirichlet boundary.

Second, the convergence rate is investigated, which is important to estimate the benefit of a finer grid.
This is a serious consideration, since for simulation in $3D$ a halved voxel size leads to a cubically increase of grid-points.
As a result, the available computing resource, such as memory, might exceed very fast.

\begin{figure}[ht]
  \centering
  \includegraphics[width=\figwidth\linewidth]{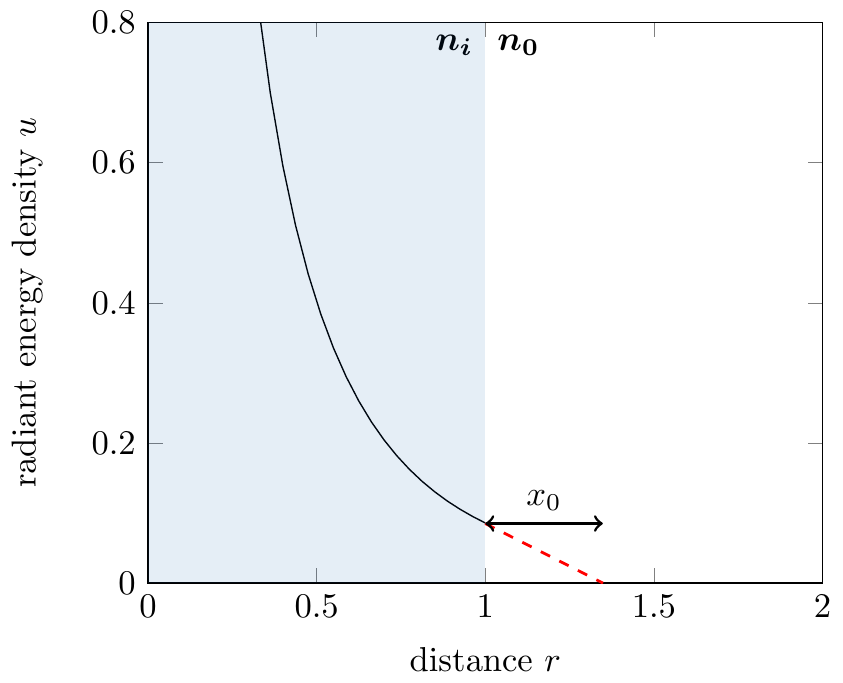}
  \caption{Solid line represents simulated data for medium with refractive index~$n_i$.
           Literature predicts length $x_0 := 2 C_R D $~\citep{wang:07}, that is given by the intersection of the tangent (red dashed line) with the x-axis.
           For the above parameters it holds $x_0 = 0.3492.$
           }
  \label{fig:analyticallength}
\end{figure}

\begin{figure}[h]
  \centering
  \includegraphics[width=\figwidth\linewidth]{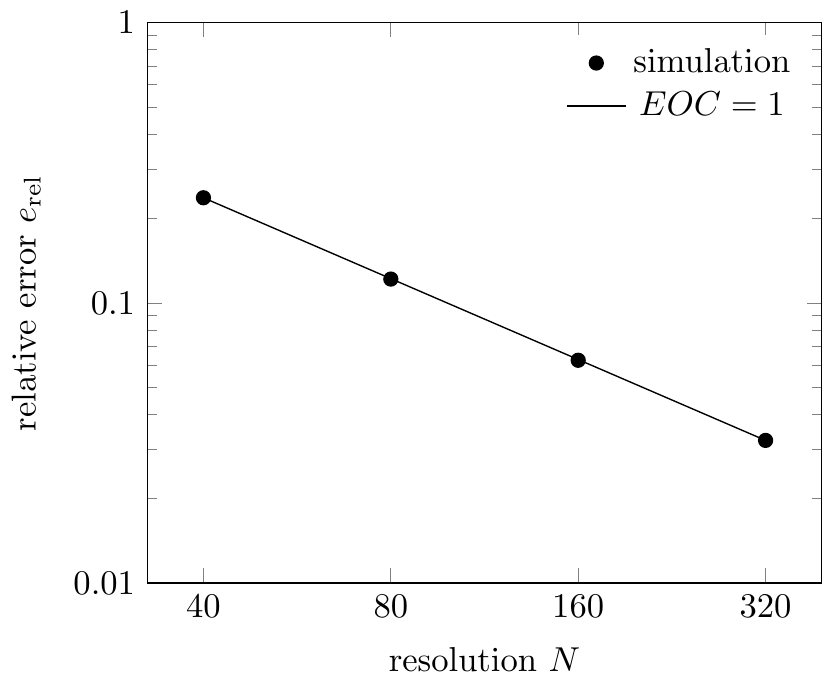}
  \caption{Relative error from simulation plotted over resolution in a log-log graph.
           An experimental order of convergence of \num{0.9588} is observed.
           Parameter choice $\sigma_a=0.5$, $\sigma_s=1.5$ and $C_R = 1.0478$.
           }
  \label{fig:analyticalLengthLogLog}
\end{figure}

Given the optical properties of the media, literature predicts a length
\begin{equation*}
x_0 = 2D C_R
\;,
\end{equation*}
after which a tangent on the boundary density intersects with the x-axis, see Fig.~\ref{fig:analyticallength} and work of Haskell et al.~\citep{haskell:94}.
It is important to note, that this validation requires the spherical geometry setup.
Here, the simulation is validated by means of computing first the density gradient by a simple two-point form and second, get the intersection with the x-axis.
The determined length~$x_{0,N}$ is then associated to the resolution of the simulation and the \emph{relative error} is introduced by
\begin{equation*}
e_{\textrm{rel},N}
=
\frac{x_0 - x_{0,N}}{x_0}
\;.
\end{equation*}
Figure~\ref{fig:analyticalLengthLogLog} depicts the relative errors for varying resolutions.
It is seen that the relative error decreases by factor of~\num{0.5} by equally refining the grid by factor of~\num{0.5}.
This means that the boundary scheme is first order convergent, which is commonly expressed by the \emph{experimental order of convergence} (\emph{EOC})
\begin{equation*}
EOC_{N,N'} = \left| \frac{\log(e_{\textrm{rel},N}) -\log(e_{\textrm{rel},N'})}{\log(\frac{1}{N}) -\log(\frac{1}{N'})} \right|
\;.
\end{equation*}
The \emph{EOC} is basically the absolute slope of the line in the log--log representation in Fig.~\ref{fig:analyticalLengthLogLog} and accounts for the speed of convergence by means of grid refinement.
Simulation data results in $EOC_{40,320}=\num{0.9588}$ and suggests that the proposed Fresnel boundary scheme is first order convergent.
A resolution of~\num{320} results in more than~\num{125e6} voxels and has been executed on a high performance cluster at the Karls\-ruhe Institute of Technology.
It is expected that the relative error can be reduced further by applying a halfway bounce-back scheme, as literature reports~\citep{zhang:12,ziegler:93}.

\subsection{Consistency to Dirichlet and Neumann boundary}
\begin{figure}[ht]
  \centering
  \includegraphics[width=\figwidth\linewidth]{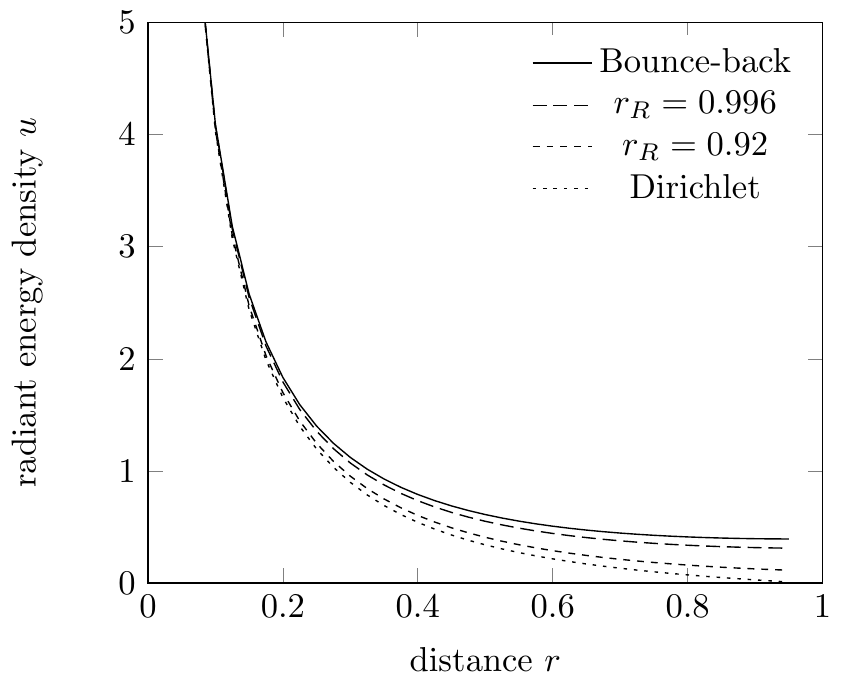}
  \caption{Proposed Robin boundary condition is consistent to zero light and zero flux boundary.}
  \label{fig:threshold}
\end{figure}

One benefit of the Fresnel boundary condition for RTLBM is the simple partial bounce-back based algorithm, where parameter $r_F \in [-1,1]$ acts similar to a threshold.
For limiting case $r_F \to 1$ the classical bounce-back scheme is obtained and hence the zero density flux (Neumann) boundary condition, from a macroscopic perspective.
Where as the limiting case $r_F \to 0$ is shown to recover zero light density (Dirichlet), see Fig.~\ref{fig:threshold}.
Both quantitative observations are also predicted from the modeling equation~\eqref{eq:robin}, which shows the consistency of the mesoscopic scheme to the macroscopic target equation.

\subsection{Application to multiple LED spots}
Simulating a light source embedded in a boundary, can be realized by two Dirichlet boundaries with different densities for the light source and a density on the outlet.
Generally, the light density at the outlet is not known, thus it is often set to zero and hence lacks on physical meaning.
In addition, the simulation is very unstable due to the jump of density.

However, the novel boundary model is capable of simulating general outlet boundaries, taking the reflectivity into account.
Applications such as multiple LED spots demonstrate the need of such boundary schemes for RTLBM.
\begin{figure}
\centering
  \subcaptionbox{For Dirichlet boundary we observe a poor penetration depth of light into the media.
    Already after short distance, the light reduced by factor of \num{10}.}
  {\includegraphics[width=\figwidth\linewidth]{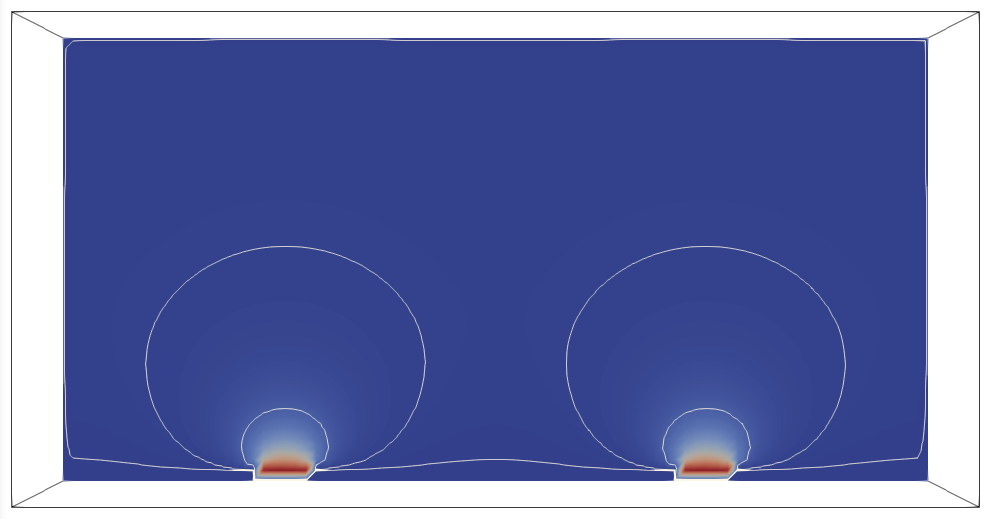}}
  \subcaptionbox{The boundary is realized by the novel Fresnel boundary.
    In general the radiant energy density in the domain is significantly higher, since there is in particular reflectivity at the bottom plate.}
  {\includegraphics[width=\figwidth\linewidth]{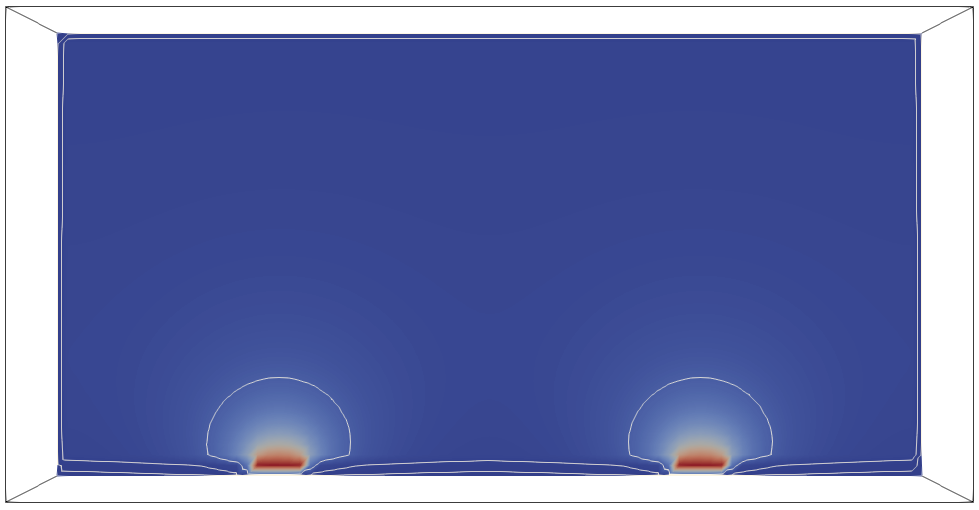}}
  \caption{Cross-section with contour lines for values \numlist{0.1; 0.01; 0.001} of radiant energy density $u$.
  Radiative source is initialized with density of \num{1}.}
\label{fig:led}
\end{figure}
The geometry setup in Figure~\ref{fig:led} is based on a cuboid with extension of $\SI{0.2}{\meter}\times\SI{0.1}{\meter}\times\SI{0.1}{\meter}$.
At the bottom plate there are two rectangular light sources ($\SI{0.01}{\meter}\times\SI{0.01}{\meter}$) that are realized by a Dirichlet boundary condition.
Other boundaries are modeled by the developed reflection boundary.

The comparison of the conservative simulation the Dirichlet boundary and the novel Fresnel boundary is shown in Fig.~\ref{fig:led}.
The Fresnel boundary is imposed for a refractive index $\nrel=0.88$, which corresponds to water-glass interface, and the volume specific parameters $\sigma_a=\SI{3}{\per\meter}$ and $\sigma_s=\SI{27}{\per\meter}$.
This setup predicts a remarkably higher light density in the volume.
This can be explained by the partially reflected radiation on the boundary, instead of fixing the light density to almost zero by~\num[scientific-notation=true]{1e-6}.

\begin{figure}
     \begin{subfigure}[b]{0.47\textwidth}
         \centering
         \includegraphics[width=\textwidth]{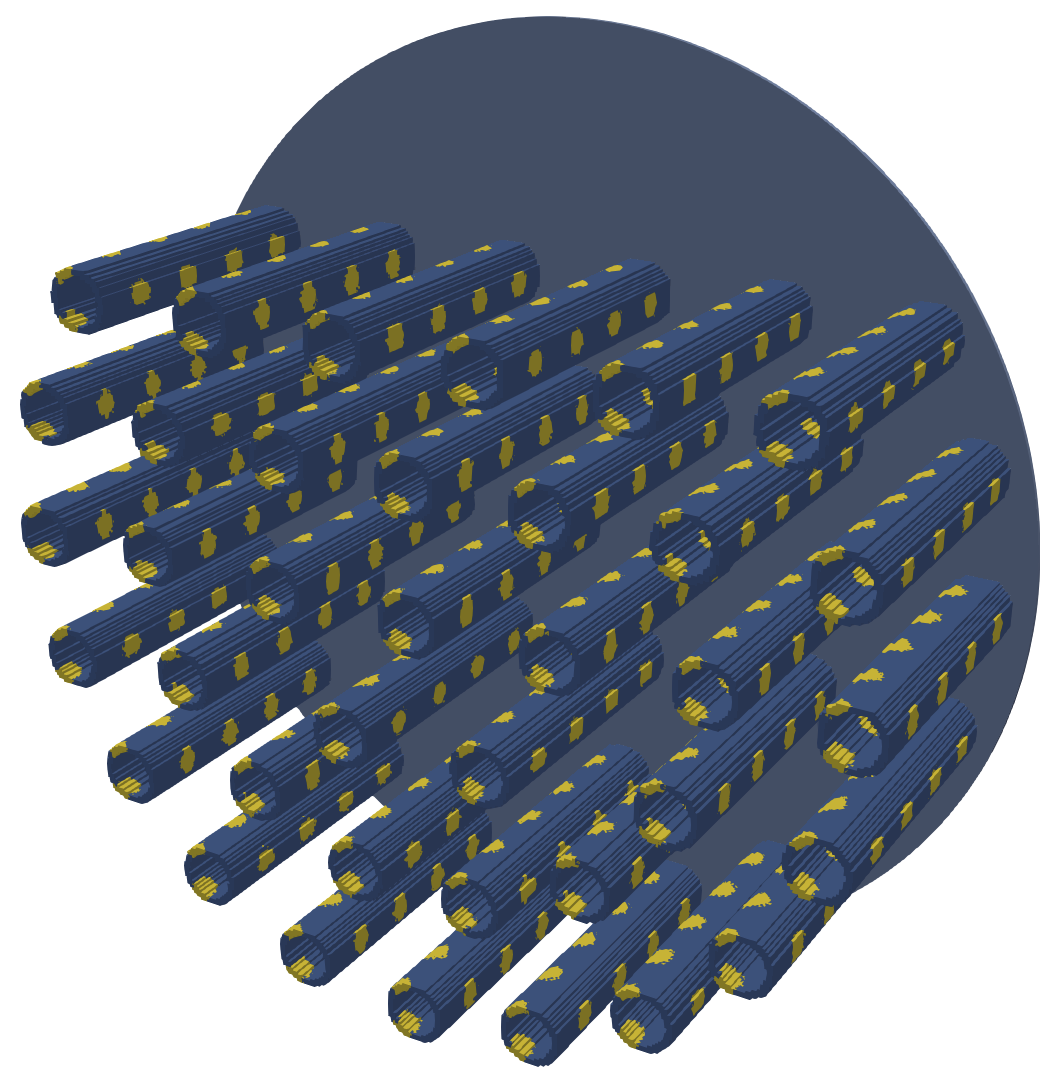}
         \caption{Geometry setup with tubular inlets equipped with multiple LED spots.}
     \end{subfigure}
\hfill
     \begin{subfigure}[b]{0.47\textwidth}
         \centering
         \includegraphics[width=\textwidth]{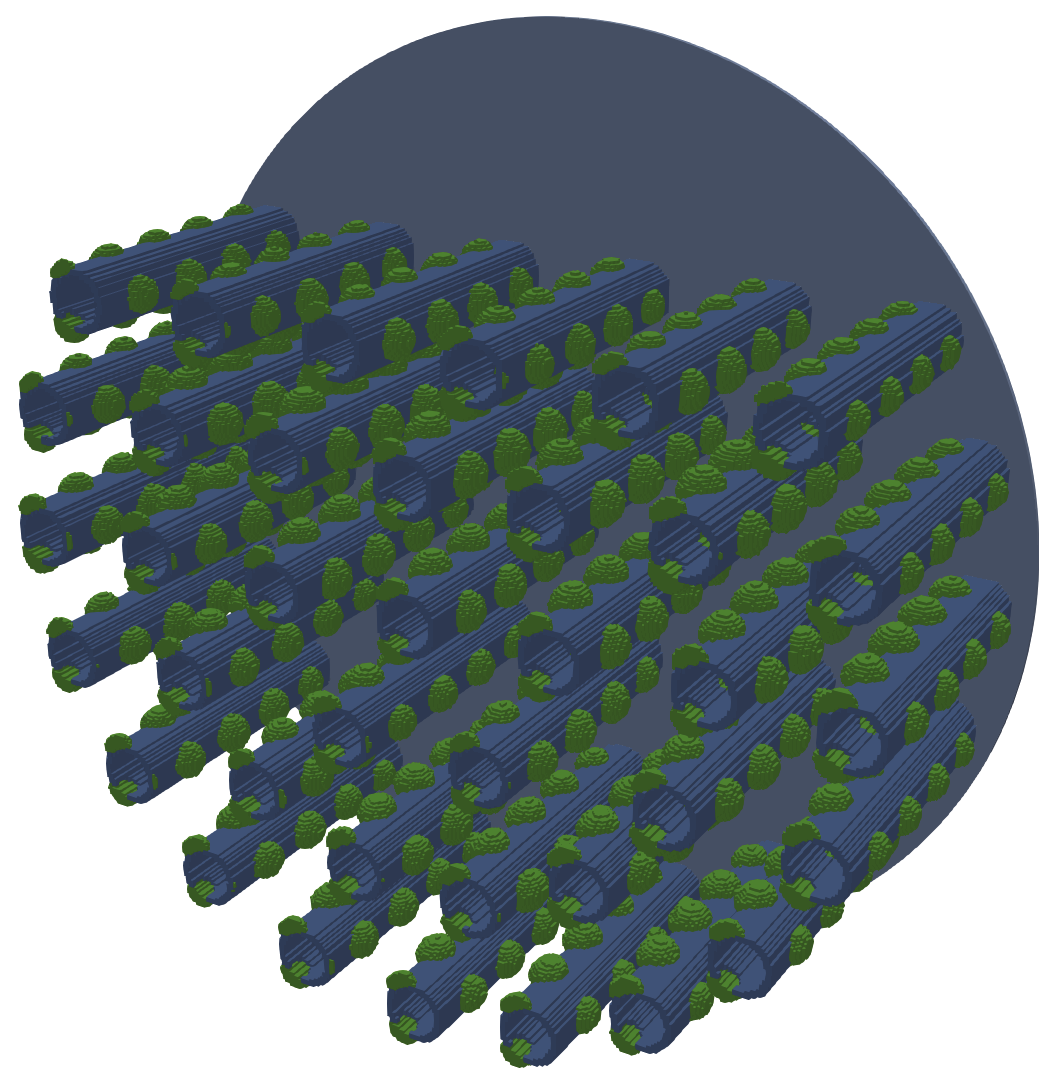}
         \caption{Light distribution for \SI{500}{\nano\meter} for a micro algae concentration of \SI{0.05}{\gram\per\liter}.}
     \end{subfigure}
     \caption{The complex geometry and the multiple LED spots require a carefully modeling of the boundary.
     Here, the blue part are outlet boundaries that are modeled by the novel reflection boundary.
     }
\label{fig:alfgine}
\end{figure}

Furthermore, the novel reflection boundary is applied to a complex geometry with curved boundaries and multiple light sources.
Here, inside a tubular reactor there are several inlets equipped with multiple LED spots, see Fig.~\ref{fig:alfgine}.
All installations inside the reactor as well as the reactor surface need to dealed with as boundaries.
Especially the boundary close to the light sources require a precise modeling, such that the reactor design can be evaluated with regard to light distribution and light inhibition can be prevented.
Thus, the physical understanding of the reflection in terms of reflection indices plays an important role for engineering applications.

\section{Conclusion}
The interaction of radiation on surfaces was investigated by the help of the Fresnel's equation.
This modeling equation depicts the partly reflected radiation on the boundary layer and allows to implement not only purely transmissive or absorbing boundaries.
Based on the Fresnel's equation, it was derived a partial bounce-back scheme for radiative transport LBM, governed only by the reflectivity of the surface.
As a consequence the benefits, of simple implementation and excellent performance, of the standard bounce-back boundary in LBM are inherited.

First, the Fresnel's equation was approximated by a Robin equation on macroscopic level, deploying the $P_1$-Approximation that assumes a highly scattering volume.
As a result the boundary reflection is then understood in terms of partial differential equations.
After that, the numerical parameters were clearly linked to modeling parameters by performing a Chapman--Enskog analysis.
In addition we showed, that the derived boundary model covers the limiting case of a transmissive ($n_i=n_o$) and purely absorbing boundary.
Finally, the numerical experiments are in very good agreement to theory and suggests, that the boundary scheme is first order convergent.

To emphasis the relevance for the proposed boundary model two different simulations of a multiple LED setups, which differ in terms of the boundary model, were shown.
The obtained results show, that the developed Fresnel boundary model implements the reflection on the surface, which leads to significant higher radiance in the domain.
This work will help to extend the usage of radiative transport LBM to situations, where the boundary effects cannot be neglected and have to be resolved physically correctly to obtain reliable results.

However, the microscopic nature and in particular, the specular reflection is not resolved exactly, due to the approximating approach.
For scattering dominated radiative transport, this model error is supposed to be small enough and the numerical model still converges~\cite{haskell:94,flock:89}.
An interesting approach are slip boundaries, already discussed in fluid flow LBM literature~\citep{succi:02, sbragaglia:05,augusto:17}, which could model specular reflection, as the re-emitted direction depends here on the incident angle.
This could be necessary in very dilute volume, where the $P_1$-Approximation is no longer valid.

\section*{Acknowledgment}
This work was performed on the computational resource ForHLR~I (ForHLR~II) funded by the Ministry of Science, Research and the Arts Baden-W\"urttemberg and DFG ("Deutsche Forschungsgemeinschaft").
This work was supported by the Deutsche Forschungsgemeinschaft (DFG) Grant No. 322739165.



\begin{thebibliography}{10}
\expandafter\ifx\csname url\endcsname\relax
  \def\url#1{\texttt{#1}}\fi
\expandafter\ifx\csname urlprefix\endcsname\relax\def\urlprefix{URL }\fi
\expandafter\ifx\csname href\endcsname\relax
  \def\href#1#2{#2} \def\path#1{#1}\fi

\bibitem{mohrhard:19}
M.~Mohrhard, G.~Th{\"a}ter, J.~Bludau, B.~Horvat, M.~J. Krause,
  Auto-vectorization friendly parallel lattice {B}oltzmann streaming scheme for
  direct addressing, Computers \& Fluids 181 (2019) 1 -- 7.
\newblock \href {https://doi.org/10.1016/j.compfluid.2019.01.001}
  {\path{doi:10.1016/j.compfluid.2019.01.001}}.

\bibitem{aidun:10}
C.~K. Aidun, J.~R. Clausen, Lattice-{B}oltzmann method for complex flows,
  Annual review of fluid mechanics 42 (2010) 439--472.
\newblock \href {https://doi.org/10.1146/annurev-fluid-121108-145519}
  {\path{doi:10.1146/annurev-fluid-121108-145519}}.

\bibitem{krueger:17}
T.~Kr{\"u}ger, H.~Kusumaatmaja, A.~Kuzmin, O.~Shardt, G.~Silva, E.~M. Viggen,
  The Lattice Boltzmann Method, Springer, 2017.

\bibitem{succi:18}
S.~Succi, S.~Succi, The lattice {B}oltzmann equation: for complex states of
  flowing matter, Oxford University Press, 2018.

\bibitem{geist:04}
R.~Geist, K.~Rasche, J.~Westall, R.~Schalkoff, {Lattice-Boltzmann Lighting},
  in: A.~Keller, H.~W. Jensen (Eds.), Eurographics Workshop on Rendering, The
  Eurographics Association, 2004, pp. 355--362.
\newblock \href {https://doi.org/10.2312/EGWR/EGSR04/355-362}
  {\path{doi:10.2312/EGWR/EGSR04/355-362}}.

\bibitem{asinari:10}
P.~Asinari, S.~C. Mishra, R.~Borchiellini, A lattice {B}oltzmann formulation
  for the analysis of radiative heat transfer problems in a participating
  medium, Numerical Heat Transfer, Part B: Fundamentals 57~(2) (2010) 126--146.
\newblock \href {https://doi.org/10.1080/10407791003613769}
  {\path{doi:10.1080/10407791003613769}}.

\bibitem{mishra:14}
S.~C. Mishra, H.~Poonia, R.~R. Vernekar, A.~K. Das, Lattice {B}oltzmann method
  applied to radiative transport analysis in a planar participating medium,
  Heat Transfer Engineering 35~(14-15) (2014) 1267--1278.
\newblock \href {https://doi.org/10.1080/01457632.2013.876806}
  {\path{doi:10.1080/01457632.2013.876806}}.

\bibitem{bindra:12}
H.~Bindra, D.~V. Patil, Radiative or neutron transport modeling using a lattice
  {B}oltzmann equation framework, Phys. Rev. E 86 (2012) 016706.
\newblock \href {https://doi.org/10.1103/PhysRevE.86.016706}
  {\path{doi:10.1103/PhysRevE.86.016706}}.

\bibitem{ma:11}
Y.~Ma, S.~Dong, H.~Tan, Lattice {B}oltzmann method for one-dimensional
  radiation transfer, Phys. Rev. E 84 (2011) 016704.
\newblock \href {https://doi.org/10.1103/PhysRevE.84.016704}
  {\path{doi:10.1103/PhysRevE.84.016704}}.

\bibitem{mcHardy:16}
C.~McHardy, T.~Horneber, C.~Rauh, New lattice {B}oltzmann method for the
  simulation of three-dimensional radiation transfer in turbid media, Opt.
  Express 24~(15) (2016) 16999--17017.
\newblock \href {https://doi.org/10.1364/OE.24.016999}
  {\path{doi:10.1364/OE.24.016999}}.

\bibitem{mink:16}
A.~Mink, G.~Th{\"a}ter, H.~Nirschl, M.~J. Krause, A 3{D} {L}attice {B}oltzmann
  method for light simulation in participating media, Journal of Computational
  Science 17~(Part 2) (2016) 431--437, discrete Simulation of Fluid Dynamics
  2015.
\newblock \href {https://doi.org/10.1016/j.jocs.2016.03.014}
  {\path{doi:10.1016/j.jocs.2016.03.014}}.

\bibitem{yi:16}
H.-L. Yi, F.-J. Yao, H.-P. Tan, Lattice {B}oltzmann model for a steady
  radiative transfer equation, Phys. Rev. E 94 (2016) 023312.
\newblock \href {https://doi.org/10.1103/PhysRevE.94.023312}
  {\path{doi:10.1103/PhysRevE.94.023312}}.

\bibitem{zhang:13}
Y.~Zhang, H.~Yi, H.~Tan, One-dimensional transient radiative transfer by
  lattice {B}oltzmann method, Opt. Express 21~(21) (2013) 24532--24549.
\newblock \href {https://doi.org/10.1364/OE.21.024532}
  {\path{doi:10.1364/OE.21.024532}}.

\bibitem{zhang:15}
Y.~Zhang, H.-L. Yi, H.-P. Tan, Lattice {B}oltzmann method for short-pulsed
  laser transport in a multi-layered medium, Journal of Quantitative
  Spectroscopy and Radiative Transfer 155 (2015) 75--89.
\newblock \href {https://doi.org/10.1016/j.jqsrt.2015.01.008}
  {\path{doi:10.1016/j.jqsrt.2015.01.008}}.

\bibitem{guo:16}
Y.~Guo, M.~Wang, Lattice {B}oltzmann modeling of phonon transport, Journal of
  Computational Physics 315 (2016) 1 -- 15.
\newblock \href {https://doi.org/10.1016/j.jcp.2016.03.041}
  {\path{doi:10.1016/j.jcp.2016.03.041}}.

\bibitem{wang:17}
Y.~Wang, L.~Yan, Y.~Ma, Lattice boltzmann solution of the transient {B}oltzmann
  transport equation in radiative and neutron transport, Phys. Rev. E 95 (2017)
  063313.
\newblock \href {https://doi.org/10.1103/PhysRevE.95.063313}
  {\path{doi:10.1103/PhysRevE.95.063313}}.

\bibitem{gairola:17}
A.~Gairola, H.~Bindra, Lattice {B}oltzmann method for solving non-equilibrium
  radiative transport problems, Annals of Nuclear Energy 99 (2017) 151 -- 156.
\newblock \href {https://doi.org/10.1016/j.anucene.2016.08.011}
  {\path{doi:10.1016/j.anucene.2016.08.011}}.

\bibitem{zhang:16}
Y.~Zhang, H.~Yi, H.~Tan, Lattice {B}oltzmann method for one-dimensional vector
  radiative transfer, Opt. Express 24~(3) (2016) 2027--2046.
\newblock \href {https://doi.org/10.1364/OE.24.002027}
  {\path{doi:10.1364/OE.24.002027}}.

\bibitem{mcHardy:18}
C.~McHardy, T.~Horneber, C.~Rauh, Spectral simulation of light propagation in
  participating media by using a lattice {B}oltzmann method for photons,
  Applied Mathematics and Computation 319 (2018) 59--70, recent Advances in
  Computing.
\newblock \href {https://doi.org/10.1016/j.amc.2017.01.045}
  {\path{doi:10.1016/j.amc.2017.01.045}}.

\bibitem{wang:07}
L.~Wang, H.~Wu, Biomedical Optics: Principles and Imaging, Wiley, 2007.

\bibitem{modest:13}
M.~F. Modest, Radiative heat transfer, Academic press, 2013.

\bibitem{wolf:95}
D.~Wolf-Gladrow, A lattice {B}oltzmann equation for diffusion, Journal of
  Statistical Physics 79~(5-6) (1995) 1023--1032.
\newblock \href {https://doi.org/10.1007/BF02181215}
  {\path{doi:10.1007/BF02181215}}.

\bibitem{patil:14}
D.~V. Patil, K.~N. Premnath, S.~Banerjee, Multigrid lattice {B}oltzmann method
  for accelerated solution of elliptic equations, Journal of Computational
  Physics 265 (2014) 172 -- 194.
\newblock \href {https://doi.org/10.1016/j.jcp.2014.01.049}
  {\path{doi:10.1016/j.jcp.2014.01.049}}.

\bibitem{cornubert:91}
R.~Cornubert, D.~d'Humières, D.~Levermore, A knudsen layer theory for lattice
  gases, Physica D: Nonlinear Phenomena 47~(1) (1991) 241 -- 259.
\newblock \href {https://doi.org/10.1016/0167-2789(91)90295-K}
  {\path{doi:10.1016/0167-2789(91)90295-K}}.

\bibitem{ziegler:93}
D.~P. Ziegler, Boundary conditions for lattice {B}oltzmann simulations, Journal
  of Statistical Physics 71~(5-6) (1993) 1171--1177.
\newblock \href {https://doi.org/10.1007/BF01049965}
  {\path{doi:10.1007/BF01049965}}.

\bibitem{haskell:94}
R.~C. Haskell, L.~O. Svaasand, T.-T. Tsay, T.-C. Feng, M.~S. McAdams, B.~J.
  Tromberg, Boundary conditions for the diffusion equation in radiative
  transfer, J. Opt. Soc. Am. A 11~(10) (1994) 2727--2741.
\newblock \href {https://doi.org/10.1364/JOSAA.11.002727}
  {\path{doi:10.1364/JOSAA.11.002727}}.

\bibitem{zhang:12}
T.~Zhang, B.~Shi, Z.~Guo, Z.~Chai, J.~Lu, General bounce-back scheme for
  concentration boundary condition in the lattice-{B}oltzmann method, Phys.
  Rev. E 85 (2012) 016701.
\newblock \href {https://doi.org/10.1103/PhysRevE.85.016701}
  {\path{doi:10.1103/PhysRevE.85.016701}}.

\bibitem{flock:89}
S.~T. Flock, M.~S. Patterson, B.~C. Wilson, D.~R. Wyman, Monte {C}arlo modeling
  of light propagation in highly scattering tissues. i. {M}odel predictions and
  comparison with diffusion theory, IEEE Transactions on Biomedical Engineering
  36~(12) (1989) 1162--1168.
\newblock \href {https://doi.org/10.1109/TBME.1989.1173624}
  {\path{doi:10.1109/TBME.1989.1173624}}.

\bibitem{succi:02}
S.~Succi, {Mesoscopic Modeling of Slip Motion at Fluid-Solid Interfaces with
  Heterogeneous Catalysis}, Phys. Rev. Lett. 89 (2002) 064502.
\newblock \href {https://doi.org/10.1103/PhysRevLett.89.064502}
  {\path{doi:10.1103/PhysRevLett.89.064502}}.

\bibitem{sbragaglia:05}
M.~Sbragaglia, S.~Succi, Analytical calculation of slip flow in lattice
  {B}oltzmann models with kinetic boundary conditions, Physics of Fluids 17~(9)
  (2005) 093602.
\newblock \href {https://doi.org/10.1063/1.2044829}
  {\path{doi:10.1063/1.2044829}}.

\bibitem{augusto:17}
L.~d. L.~X. Augusto, J.~Ross-Jones, G.~C. Lopes, P.~Tronville, J.~A.~S.
  Gon{\c{c}}alves, M.~R{\"a}dle, M.~J. Krause, Microfiber filter performance
  prediction using a lattice {B}oltzmann method, Commun Comput Phys 23 (2018)
  910--931.
\newblock \href {https://doi.org/10.4208/cicp.OA-2016-0180}
  {\path{doi:10.4208/cicp.OA-2016-0180}}.

\bibitem{he:00}
X.~He, N.~Li, B.~Goldstein, {Lattice Boltzmann Simulation of
  Diffusion-Convection Systems with Surface Chemical Reaction}, Molecular
  Simulation 25~(3-4) (2000) 145--156.
\newblock \href {https://doi.org/10.1080/08927020008044120}
  {\path{doi:10.1080/08927020008044120}}.

\bibitem{kang:07}
Q.~Kang, P.~C. Lichtner, D.~Zhang, {An improved lattice Boltzmann model for
  multicomponent reactive transport in porous media at the pore scale}, Water
  Resources Research 43~(12) (2007).
\newblock \href {https://doi.org/10.1029/2006WR005551}
  {\path{doi:10.1029/2006WR005551}}.

\bibitem{chai:08}
Z.~Chai, B.~Shi, A novel lattice {B}oltzmann model for the {P}oisson equation,
  Applied Mathematical Modelling 32~(10) (2008) 2050 -- 2058.
\newblock \href {https://doi.org/10.1016/j.apm.2007.06.033}
  {\path{doi:10.1016/j.apm.2007.06.033}}.

\end{thebibliography}

\appendix
\section{Chapman--Enskog analysis}
\label{sec:ce}

Based on the Chapman--Enskog expansion, the mesoscopic formulation is related to the target Robin equation.
This approach ensures that the simulation parameter on mesoscopic level are properly linked to the macroscopic parameters in the Robin equation.
Recall, that the stream and collide equation~\eqref{eq:rtlbm} for the highly scattering volume is given by
\begin{equation*}
f_i(t+\triangle t, \xx + \sss_i \triangle t)
=
f_i(t,\xx)
-
\frac{\triangle t}{\tau}
( f_i -f^{eq}_i ) (t,\xx)
-
\triangle t^2 \, \eta \, f_i(t,\xx)
\;,
\end{equation*}
for $\eta := \frac{3\,\sigma_a^\star(\sigma_a^\star+\sigma_s^\star)}{8}$ and discretization parameter $\triangle t$.
Firstly, a Taylor series expansion up to second order in $\triangle t$ of the left hand side reads
\begin{multline*}
f_i(t +\triangle t,\xx+\sss_i \triangle t)
=
f_i(t,\xx)
+
\triangle t
\Big(
  \partial_t + \sss_i \cdot \nabla
\Big)
f_i(t,\xx)
+
\frac{\triangle t^2}{2}
\Big(
  \partial_t + \sss_i \cdot \nabla
\Big)^2
f_i(t,\xx)
.
\end{multline*}
Substituting this expression into the stream and collide equation, it holds
\begin{equation}
\label{eq:afterTaylor}
\Big(
  \partial_t + \sss_i \cdot \nabla
\Big)
f_i
+
\frac{\triangle t}{2}
\Big(
  \partial_t + \sss_i \cdot \nabla
\Big)^2
f_i
=
\frac{1}{\tau}
( f_i -f^{eq}_i )
-
\triangle t \, \eta \, f_i
\;.
\end{equation}
By introducing a small, positive scaling parameter $\varepsilon^2 = \triangle t$, the time and space derivation can be rewritten
\begin{equation}
\label{eq:scaling}
\partial_t  \leadsto \varepsilon^2 \partial_t\;,  \quad \quad
\nabla      \leadsto \varepsilon   \nabla
\;.
\end{equation}
In addition the particle density function $f_i$ is expanded by
\begin{equation}
\label{eq:expansionCE}
f_i
=
f_i^{(0)} + \varepsilon f_i^{(1)} + \varepsilon^2 f_i^{(2)} + \ldots
\; .
\end{equation}

Using the scaling~\eqref{eq:scaling} and the expansion~\eqref{eq:expansionCE}, equation~\eqref{eq:afterTaylor} can be rewritten in the consecutive orders of parameter $\varepsilon$ as
\begin{align}
\label{eq:eps0}
\varepsilon^0&:
f_i^{(0)} = f_i^{eq}
\;,
\\
\label{eq:eps1}
\varepsilon^1&:
\sss_i \cdot \nabla
  f_i^{(0)}
=
-\frac{1}{\tau}
  f_i^{(1)}
\;,
\\
\label{eq:eps2}
\varepsilon^2&:
\partial_t f_i^{(0)}
+
\sss_i \cdot \nabla f_i^{(1)}
=
-
\frac{1}{\tau} f_i^{(2)}
-
\eta \, f_i^{(0)}
\;.
\end{align}
From~\eqref{eq:expansionCE} and~\eqref{eq:eps1}, it follows
\begin{equation*}
\sss_i \cdot \nabla f_i^{eq}
=
-\frac{1}{\tau \varepsilon}
\left(
f_i -f_i^{eq}
\right)
\;,
\end{equation*}
or equivalently
\begin{equation}
\label{eq:robinCE}
w_i\, \sss_i \cdot \nabla u
=
-\frac{1}{\tau \varepsilon}
\left(
f_i -w_i u
\right)
\;,
\end{equation}
according to the definition of the equilibrium function $f_i^{eq} = w_i u$.

To find a closure for the system~\eqref{eq:eps0}, \eqref{eq:eps1} and~\eqref{eq:eps2}, the discrete version of the Robin boundary condition~\eqref{eq:robin}
\begin{equation*}
w_i\, \sss_i \cdot \nabla u = - \frac{w_i}{2C_R D}\, u
\end{equation*}
is substituted by~\eqref{eq:robinCE} leading to
\begin{equation*}
w_i u
=
\frac{f_i}{ 1+\frac{\tau \varepsilon}{2C_R D}}
\;.
\end{equation*}
For diffusion coefficient $D$ in meter and diffusive  scaling $\Delta t \propto  \Delta x^2$.
Further, the closure assumes that the incoming non-equilibrium parts equal the outgoing ones in the opposite direction~\citep{he:00, kang:07}.
This relation reads
\begin{equation*}
f_{\bar{i}} - f_{\bar{i}}^{eq}
=
-\left(
f_i - f_i^{eq}
\right)
\end{equation*}
and is rewritten to
\begin{equation}
\label{eq:reflect}
f_{\bar{i}}
=
\left(
\frac{2}{1+\frac{\varepsilon}{2C_R D}} -1
\right)
f_i
\;.
\end{equation}
Note, in RTLBM it is common to set~$\tau$ equals one~\citep{mink:16,chai:08}.
The definition of~$f^{eq}_i$ and the underlying stencil~$D3Q7$ yields in $f_i^{eq} = f_j^{eq}$ for all $i,j\neq0$.
The derived equation can be interpreted as the discrete version of the boundary equation~\eqref{eq:mesoscopicrobin}.
Equation~\eqref{eq:reflect} is a general partial bounce\hbox{-}back formula and is derived through the performed Chapman--Enskog expansion.

\end{document}